# MonLAD: Money Laundering Agents Detection in Transaction Streams*

Xiaobing Sun[1,2 †], Wenjie Feng[3 †], Shenghua Liu[1,2], Yuyang Xie[5], Siddharth Bhatia[4], Bryan Hooi[4],
Wenhan Wang[6], Xueqi Cheng[1,2]

[1]Insititute of Computing Technology, Chinese Academy of Sciences, Beijing, China
[2]University of Chinese Academy of Sciences, Beijing, China
[3] Institute of Data Science, National University of Singapore   [4]National University of Singapore, Singapore
[5]Tsinghua University, Beijing, China   [6]Tencent Technology Co. Ltd, ShenZhen, China
sunxiaobing21@mails.ucas.ac.cn,wenchiehfeng.us@gmail.com,liushenghua@ict.ac.cn

## ABSTRACT

Given a stream of money transactions between accounts in a bank, how can we accurately detect money laundering agent accounts and suspected behaviors in real-time? Money laundering agents try to hide the origin of illegally obtained money by dispersive multiple small transactions and evade detection by smart strategies. Therefore, it is challenging to accurately catch such fraudsters in an unsupervised manner. Existing approaches do not consider the characteristics of those agent accounts and are not suitable to the streaming settings. Therefore, we propose MonLAD and MonLAD-W to detect money laundering agent accounts in a transaction stream by keeping track of their residuals and other features; we devise AnoScore algorithm to find anomalies based on the robust measure of statistical deviation. Experimental results show that MonLAD outperforms the state-of-the-art baselines on real-world data and finds various suspicious behavior patterns of money laundering. Additionally, several detected suspected accounts have been manually-verified as agents in real money laundering scenario.

## CCS CONCEPTS

• **Information systems** → **Electronic commerce**; • **Theory of computation** → **Streaming, sublinear and near linear time algorithms**; • **General and reference** → **Experimentation**.

## KEYWORDS

Anomaly detection, Money laundering, Stream algorithm.

*† Both authors contributed equally to the paper. Corresponding author: S. Liu.

## 1 INTRODUCTION

Money laundering is the process aiming at hiding the origin of illegally obtained money. The estimated amount of money laundered per year is $2 - 5\%$ of the global GDP, or $800 billion - $2 trillion US dollars and even a lower estimate is underlining the severity of the issue [42]. Source money could come from illegal dealing of commodities, drug trafficking, smuggling, and other criminal activities. Once connected with organized crime and terrorist financing, this will extensively damage the reputation of financial institutions and threaten public security [22].

Thus, it raises the following question: given a stream of transactions for money transfers between accounts, how can we effectively detect money laundering agent accounts in real-time? Money laundering is particularly hard to detect because fraudsters intentionally evade detection via innovative mechanisms, e.g., by funneling money through multiple accounts, mixing them with legitimate transactions, and making small transfers that fall just below the reporting thresholds [14]. Quick and accurate detection of such suspected accounts and behavior patterns in real-time transactions is the main challenge.

To transfer large amounts of money while keeping each transaction below a "safe" level, the agent accounts must make frequent incoming connections from the source accounts (or agents) and immediate or periodic outgoing connections to the target accounts. These agent accounts exhibit suspiciously fast incoming and outgoing transfer behaviors. Fig. 1 depicts an example of money laundering agents about their transfer structures and behavior series.

Most existing money laundering detection approaches [15, 17, 19, 24, 29, 37] are designed for static records and cannot adapt to streaming scenario. Methods that detect anomalies based on dynamic graphs [3, 7, 8, 34] or outlier detection [4, 9, 28, 40, 41] do not consider the characteristics of money-laundering behavior, resulting in the inferior detection accuracy or not applicable.

In this paper, we propose MonLAD, a scalable sketching algorithm for depicting the behavior of money laundering agent accounts in a transaction stream; it keeps track of the account residual and computes some key statistical features to summarize their behavior; MonLAD-W accurately fits complex patterns in a sliding window manner. We devise AnoScore algorithm to detect anomalies based on the robust statistical deviation, which is theoretically founded and explainable. Experiments on the real-world data show that our method can fast and accurately detect various patterns for money laundering (see Fig. 2(a)-2(b)) and find many suspicious agent accounts (e.g., Fig. 1), some of which are manually verified.

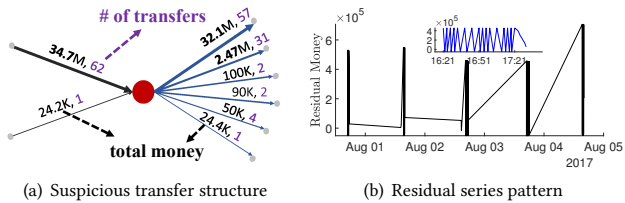

(a) Suspicious transfer structure    (b) Residual series pattern

Figure 1: Money laundering agent behavior pattern example. (a) the connections of incoming and outgoing transfers of a detected suspicious agent (red node, w.r.t. $p_3$ in Fig. 2(a)). (b) the residual series over time of the suspicious agent account, where it shows many suspicious periodical and intensive spikes for receiving money and transferring all out at once. The top subfigure zooms into part of the time.

We make all of our algorithm and experimental code and the majority of the data for the experiments available[1].

Summary of Contributions.
- **Streaming algorithm:** We propose a novel approach MonLAD to detect money laundering agent accounts in a transaction stream, it is able to fast answer the detection query at *any time* based on the statistical features (see Figure 2(a)).
- **Explainable and Flexible:** The proposed AnoScore utilizes statistical deviations as the anomaly score instead of a "black box". Our method can easily adapt to and incorporate different scores from other outlier detection methods.
- **Effectiveness:** Our method outperforms the state-of-the-art baselines on the real-world bank data, it detects various suspicious behavior patterns (e.g., Figure 1), including manually verified fraudulent accounts and periodical patterns.
- **Scalability:** MonLAD is scalable, with linear time complexity in the number of edges of the stream (see Figure 2(c)).

## 2 RELATED WORK

In this section, we review relevant approaches about anomaly and outlier detection, money laundering pattern detection.

**(Semi-) Supervised learning methods:** To involve more attributes and handle high-dimensional data, machine learning models such as SVM [39], decision trees [43], and neural networks [20] are applied to money laundering. [38] uses representation learning techniques to utilize the information contained in graphs. [32] combines network analysis to detect the groups of money laundering activities. Although these algorithms detect money laundering activities in supervised or semi-supervised manners, they suffer from imbalanced class and lacking of adaptability. We focus on detecting money laundering activities in an unsupervised fashion.

**(Streaming-) Graph-based methods:** Given as input a stream of edges over time, Goutlier [1] scores the likelihood of each edge in the stream based on a structural reservoir sample of edges. SedanSpot [7] and [27] measure edge anomalousness in the stream based on its prior occurrence, preferential attachment and mutual neighbors (homophily). Spotlight [8] detects sudden appearance of many unexpected edges. [23] applies only when multiple graphs with typed nodes and edges evolve simultaneously. Midas [3] identifies micro-cluster based anomalies or suddenly arriving groups of suspiciously similar edges. Nevertheless, those streaming methods do not take into account the characteristics of money laundering. [19] constructed a classifier based on a set of mined rules to detect suspicious transactions in a data stream. However, these traditional rule based algorithms highly rely on domain knowledge and also are easy to be evaded by fraudsters. FlowScope [17] and AutoAudit [16] detect the flow of money laundering in a multipartite graph, however, they cannot handle streaming edges and fail to capture a variety of different behavior patterns well.

Table 1 compares MonLAD and other related methods for the problem of money laundering pattern detection in streams.

Table 1: Comparison of MonLAD and relevant approaches.

| | FlowScope [17] | Midas-R [3] | DenseAlert [34] | Spotlight [8] | SedanSpot [7] | AugSplicing [44] | AutoAudit [16] | MonLAD |
|---|---|---|---|---|---|---|---|---|
| **Money laundering** | ✓ | | | | | | ✓ | ✓ |
| **Streaming pattern** | | ✓ | ✓ | ✓ | ✓ | ✓ | | ✓ |
| **Statistical deviation** | | ✓ | | ✓ | ✓ | | | ✓ |
| **Real-valued feature** | ✓ | | ✓ | ✓ | ✓ | ✓ | ✓ | ✓ |

## 3 PROBLEM FORMULATION

Here, we summarize three key traits of typical money laundering transfers and then formally define our problem. We use "fan-in" and "fan-out" to refer to the money transfer into and out of an account (especially an agent in intermediate accounts $M$) as a metaphor for fan-like incoming / outgoing connections in social networks.

### 3.1 Key Traits

**Trait 1 (Fast Fan-in and Fan-out).** *"Dirty money" will be divided into multiple parts and transferred from sources to destinations; these small-amount but multiple transfers flowing through $M$ are usually completed within a short period.*

Thus, by controlling the amount per transfer and the number of transactions, fraudsters evade detection by manipulating a large amount of money. The sooner these transactions complete, the more the fraudsters gain and lower the risk.

**Trait 2 (Frequently Balancing).** *Agent accounts crave to frequently reach a balanced state by fan-outs (transfer out to other agents or targets), upon receiving all the money or reaching a volume threshold of multiple fan-ins from the sources.*

The remaining money in the agent accounts will be subject to the risk of being detected or frozen, particularly for large amounts. Therefore, fraudsters tend to remain for as little time as possible unless they can empty the account. The remaining amount is then also be used as camouflage by some smart fraudsters.

**Trait 3 (High Throughput).** *To process a large amount of money in a short period with a limited number of agent accounts in*

---
[1]https://github.com/BGT-M/MonLAD.

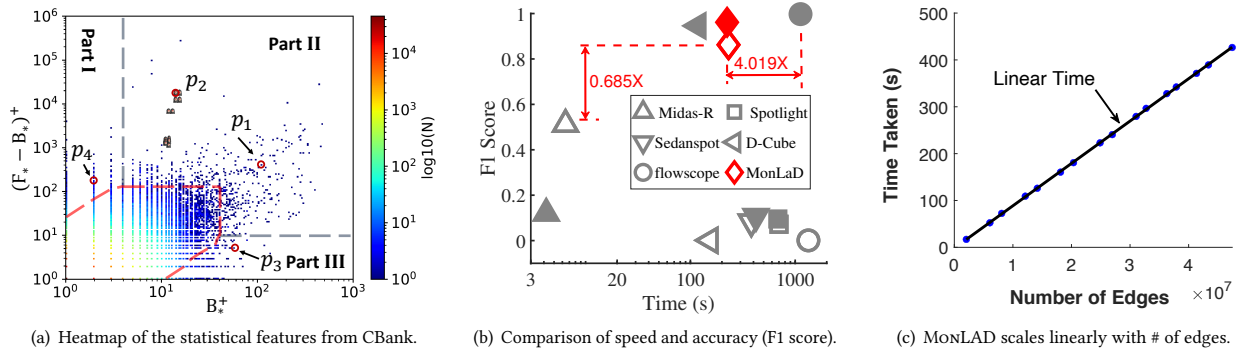

(a) Heatmap of the statistical features from CBank.  (b) Comparison of speed and accuracy (F1 score).  (c) MonLAD scales linearly with # of edges.

Figure 2: Performance of proposed MonLAD on real data (CBank). (a) Heatmap of the statistical features used in MonLAD, where criminal-icon labels 12 manually-verified accounts involved in money laundering and other outliers also exhibit obvious suspicious behavior pattern. (b) MonLAD outperforms the baselines in two different settings with empty and filled symbols for each method. (c) MonLAD runs in time linear to the number of edges in a stream.

$\mathcal{M}$, achieving more balanced counts or transferring a high amount each time is required.

Accounting for limited accessible resources (e.g., the number of agent accounts) and time constraints, fraudsters will multiplex accounts in $\mathcal{M}$ to conduct the same behavior for reducing the cost and achieving "economy of scale".

The distinctive behaviors of agents involved in money laundering can be characterized with the above traits, which distinguishes from the normal. As opposed to methods detecting anomaly patterns in static data, we spot the money laundering anomaly patterns from a large-scale transaction stream and detect the agent accounts.

### 3.2 Problem Definition

Let $\mathcal{G} = (\mathcal{V}, \mathcal{E}, \mathcal{W}, \mathcal{T})$ be a time-evolving directed graph for the money transfers, $\mathcal{E} = \{e_1, e_2, \cdots\}$ denotes a stream of edges, and the vertex set $\mathcal{V}$ represents the accounts. Each arriving edge is a tuple $e = (u, v, w_{u,v}, t)$ consisting of a source vertex $u$, a destination vertex $v$, a weight $w_{u,v} \in \mathcal{W} \in \mathbb{Z}^+$, and a time of occurrence $t \in \mathcal{T}$ at which the edge is added to $\mathcal{G}$. We use the term vertex and account interchangeably throughout the paper in terms of our scenario.

We treat $\mathcal{G}$ as a multi-graph, i.e., edges can be created multiple times between the same pair of vertices. Moreover, we do not assume that vertices $\mathcal{V}$ are known a prior in the edge stream. Table 2 gives the complete list of symbols we use.

The money laundering agents detection problem is defined as

PROBLEM 1 (MONEY LAUNDERING AGENTS DETECTION IN STREAM).
**Given** a stream of transfer records $\{e_1, e_2, \cdots\}$, where each item $e = (u, v, w_{u,v}, t)$ denotes a money transfer with an amount of $w_{u,v} > 0$ occurring at time $t$ from the account $u$ to the account $v$;
  - **Find** the group of most suspicious agent accounts $\mathcal{M}$;
  - **such that** each of accounts in $\mathcal{M}$ satisfies the Traits 1-3.

## 4 PROPOSED METHOD

In this section, we first define some key statistics to recognize the patterns of agent accounts, then propose MonLAD to sketch the statistics in a stream. Finally, we design ANOSCORE, an intuitive way to detect the suspicious accounts in money laundering setting.

Table 2: Symbols and Definitions.

| Symbol | Definition |
| --- | --- |
| $s_u^t$ | Balanced state of the account $u$ at time $t$ |
| $R_u(t)$ | Residual of the account $u$ until $t$ |
| $B_u(t)$ | the number of times of balance that the account $u$ achieves until $t$ |
| $F_u(t)$ | the total number of effective fan-ins of $u$ within balance until $t$ |
| $\delta_{up}, \delta_{down}$ | minimum thresholds for an effective fan-in and fan-out |
| $\min_u^t, \max_u^t$ | minimum and maximum residual of $u$ at $t$ during reaching balance |

### 4.1 Anomalous Balance Patterns

We define some statistics to depict the behavior of agent accounts in an edge stream.

**Weighted in-degree and out-degree:** Given time $t$, we define the weighted in-degree and out-degree[2] of the account $u$ until $t$ as

$$d_u^+(t) = \sum_{v \in \mathcal{V}} \sum_{t_i \in e(v,u,w_{v,u},t_i)}^{t} w_{v,u}; \quad d_u^-(t) = \sum_{v \in \mathcal{V}} \sum_{t_i \in e(u,v,w_{u,v},t_i)}^{t} w_{u,v}.$$

**Residual:** We define the residual as the difference between the weighted in-degree and out-degree of the account $u$ until $t$,

$$R_u(t) = d_u^+(t) - d_u^-(t) \qquad (1)$$

Note that the residual might also be negative, since the account state (initial residual) is unknown at $t_0$. $R_u(t)$ only indicates the residual after the first transaction of the account $u$ occurred during the observation period. Thus, we define two key concepts as follows:

*Definition 1 (Reaching Balanced State).* An account $u$ reaches a balanced state at $t$ (denoted as $s_u^t = 0$) after a fan-out transfer,

---
[2]They are equal to the total transfer-in and transfer-out amount until $t$ respectively.

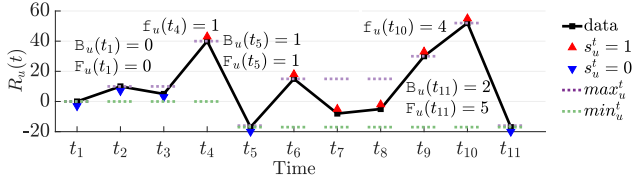

Figure 3: The residual series of the account $u$ over time, as a running example of MonLAD with $\delta_{up} = \delta_{down} = 20, \epsilon = 3$.

leading its residual to being no more than the minimum of the residual between the most recent time $\tau$ and $t$, where $\tau = 0$ or $s_u^\tau = 0$, while not balanced at $t - 1$ (i.e., $s_u^{t-1} = 1$) [3].

An account *reaches a balanced state* when the money transferred into it is finally transferred out. So we keep track of the minimum of the residual to find those balanced states. Let $B_u(t)$ be *the number of times of balanced state* that $u \in \mathcal{V}$ reaches until time $t$.

*Definition 2 (Effective Fan-in).* A money transfer into an account $u$ is called an effective fan-in if and only if
- the accumulated money transferred into $u$ (the total fan-in) reaches a threshold [4] since the last balanced state;
- all accumulated money is finally transferred out to reach the next balanced state.

In money laundering scenario, we consider the number of incoming transfers after an *effective fan-in* (inclusive) until reaching a balanced state. We use $f_u(t)$ to denote the number of effective fan-ins of the account $u$ achieves at time $t$ since the last balanced state ($u$ may (re-) reach a balanced state depending on the residual afterward), and $F_u(t)$ as the accumulative $f_u(\cdot)$ from the initial to the last balanced state before $t$.

Let $\delta_{up}$ be the minimum threshold for an effective fan-in and $\delta_{down}$ be the minimum threshold for an fan-out to be a balanced state. Consider a certain balanced state of the account $u$ from $t_0$ (the initial time or the end of the previous balance) to $t_k$, for any $t \in \{t_0, \cdots, t_k\}$, $\max_u^t$ ($\min_u^t$) denotes the maximum (minimum) residual during the process, which is traced by $s_u^t$, i.e. $s_u^t = 0$ if $t \in \{t_0, t_k\}$ otherwise $s_u^t = 1$.

Example 1 illustrates an explanation based on Figure 3, which shows the change the residual $R_u(t)$ of $u$ in $[t_1, t_{11}]$. Note that the residual becomes negative at $t \in \{t_5, t_7, t_8, t_{11}\}$ due to the unknown $R_u(t_0)$, while it must hold that $R_u(t_0) \geq \max\{|R_u(t_5)|, |R_u(t_{11})|\}$.

EXAMPLE 1 (CONCEPT EXAMPLE). *The changes of residual $R_u(t)$ in Fig. 3 come from 10 transfers, i.e., 4 fan-outs (at $t_3, t_5, t_7$, and $t_{11}$) and 6 fan-ins (at other times); it reaches balanced state at $t_5$ and $t_{11}$ according to the definition, i.e., $s_u^{t_5} = s_u^{t_{11}} = 0$. The residual achieves a new minimum at $t_5$ (i.e. $\min_u^{t_5}$) since $t_1$. Assuming $\delta_{up} = \delta_{down} = 20$, $f_u(t_4) = 1$ due to the effective fan-in at $t_4$; $F_u(t_5) = 1$ and $B_u(t_5) = 1$ since $u$ reaches balanced state only once from $t_1$ to $t_5$; $f_u(t_{10}) = 4$ since the effective fan-in at $t_6$ and there are 3 fan-ins after $t_6$; $F_u(t_{11}) = F_u(t_5) + f_u(t_{10}) = 5$ and $B_u(t_{11}) = 2$ due to the second balanced state at $t_{11}$. In addition, $\max_u^{t_2} = \max_u^{t_3} = R_u(t_2)$, $\max_u^{t_4} = \max_u^{t_5} = R_u(t_4)$; $\min_u^{t_2} = \min_u^{t_3} = \min_u^{t_4} = R_u(t_1)$. We can get other relevant values (states) after $t_5$ similarly.*

---
[3] Here '1' means waiting for balanced state and '0' refers to the start (end) of balance.
[4] Here the threshold is used to denoise some small transfers in and out of an account.

Note that $F_u(t) \geq B_u(t)$ always holds since being in the balanced state contains at least a fan-in (including effective fan-in) at any time. Regarding the money laundering scenario, some suspicious behavior patterns corresponding to different types of anomalies [30], including transfers in illegal exchange, illegal gambling, drug trafficking, etc. It can be derived as follows:
- **P1.** $F_u(t) \approx B_u(t)$: one-time fan-in and immediate fan-out;
- **P2.** $F_u(t) \gg B_u(t)$ & *a small* $B_u(t)$: dispersive fan-ins and few fan-outs.
- **P3.** $F_u(t) > B_u(t)$ & *a large* $B_u(t)$: dispersive fan-ins and fan-outs;

### 4.2 Our MonLAD Algorithm

$B_u(t)$ and $F_u(t)$ are core features to describe the behavior of the account $u$ in a transfer stream; the other statistics, including $f_u(t)$, $s_u^t$, $\min_u^t$, and $\max_u^t$, are auxiliaries; they will constantly be updated as the transaction arrives. So, we design the following rules to update and count. Alg. 1 gives the high-level pseudo-code of our method MonLAD.

When an edge $e = (u, v, w_{u,v}, t)$ arrives, we need to determine whether the source vertex $u$ reaches a balanced state due to the fan-out and whether the target vertex $v$ is waiting to reach a balanced state or will start a new process just after the previous balance. The states $s_u^t$ and $s_v^t$ will be updated according to Eq. (2) and Eq. (3) respectively, where the contribution of $w_{u,v}$ to the residuals $R_u(t)$ and $R_v(t)$ has been counted by updating $d_u^-(t), d_v^+(t)$,

$$s_u^t = \begin{cases} 0 & \text{if } \max_u^{t-1} - R_u(t) > \delta_{down} \ \& \ R_u(t) \leq \min_u^{t-1} + \epsilon, \\ s_u^{t-1} & \text{otherwise.} \end{cases} \quad (2)$$

This means that $s_u^t$ will be in a balanced state if a large enough fan-out $e$ almost clears the residual $R_u(t)$ (close to $\min_u^{t-1}$), otherwise the previous state holds. Here, we introduce a small residual $\epsilon > 0$ to nullify the fraudsters' attempts to evade detection by keeping a low balance. To be more precise, $\epsilon$ is the minimum cost of fraudsters and also the maximum tolerance of the detector.

$$s_v^t = \begin{cases} 1 & \text{if } R_v(t) - \min_v^{t-1} > \delta_{up}, \\ s_v^{t-1} & \text{otherwise.} \end{cases} \quad (3)$$

that is, $s_v^t$ is waiting to reach a balanced state if the residual $R_v(t)$ is still at least $\delta_{up}$ greater than the previous minimum $\min_v^{t-1}$, otherwise the previous state holds.

For the target vertex $v$, $f_v(t)$ will be updated if Eq. (4) satisfies $s_v^t = 1$ (as it is or due to the fan-in $e$). For the source vertex $u$, if it reaches a balanced state due to this fan-out, i.e. $s_u^t = 0$ & $s_u^{t-1} = 1$, then $f_u(t)$ will be added onto $F_u(t)$ as Eq. (5) shows and then be reset to 0, $B_u(t)$ will also be updated as Eq. (6) shows. Otherwise, they will keep the same as the previous state.

$$f_v(t) = \begin{cases} f_v(t-1) + 1 & \text{if } s_v^t = 1, \\ f_v(t-1) & \text{otherwise.} \end{cases} \quad (4)$$

$$F_u(t) = \begin{cases} F_u(t-1) + f_u(t) & \text{if } s_u^t = 0 \ \& \ s_u^{t-1} = 1, \\ F_u(t-1) & \text{otherwise.} \end{cases} \quad (5)$$

$$B_u(t) = \begin{cases} B_u(t-1) + 1 & \text{if } s_u^t = 0 \ \& \ s_u^{t-1} = 1, \\ B_u(t-1) & \text{otherwise.} \end{cases} \quad (6)$$

**Algorithm 1** MONLAD: Statistical features in stream

**Input:** Stream of edges over time; thresholds $\delta_{up}, \delta_{down}$, and $\epsilon$.
**Output:** Statistical features per vertex.
1: **while** new edge $e = (u, v, w_{u,v}, t)$ is received **do**
2:    **if** $u$ ($v$) never appeared **then** ▷ *Initialization for newcomers*
3:       INITIALIZE($u$ ($v$), $t - 1$)    ▷ *For vertex u or v*
4:    $R_u(t) = R_u(t) - w_{u,v}$;    $R_v(t) = R_v(t) + w_{u,v}$
     ▷ *Update 'state' for source u and destination v resp.*
5:    update $s^t_u, s^t_v$ based on Eqns. (2) and (3) respectively
6:    sequentially update $f_v(t), F_u(t), B_u(t)$ with Eqns. (4)-(6)
7:    **if** $s^t_u = 0 \& s^{t-1}_u = 1$ **then**
8:       $f_u(t) = 0$              ▷ *Reset $f_u(t)$*
9:    update $\min^t_u$ and $\max^t_v$ based on Eqns. (7)-(8)
10:    **output** $B_u(t), F_u(t), B_v(t), F_v(t)$.
11: **procedure** INITIALIZE($u$, $t$)    ▷ *Vertex specific variables*
12:    $R_u(t) \leftarrow 0, B_u(t) \leftarrow 0, s^t_u \leftarrow 0, f_u(t) \leftarrow 0$;
13:    $\min^t_u \leftarrow 0, \max^t_u \leftarrow 0$

Afterward, for the vertex $u$ with the fan-out $e$, $\min^t_u$ will be updated by the current residual $R_u(t)$ if it reaches balanced state or $R_u(t)$ is less than $\min^{t-1}_u$; for the vertex $v$ with the fan-in $e$, $\max^t_v$ will be updated by the residual $R_v(t)$ if $v$ starts a new process and waits for reaching a balanced state ($s^t_v = 1 \& s^{t-1}_v = 0$) or $R_v(t)$ is greater than $\max^{t-1}_v$. Eq. (7) and (8) depict the above update rules.

$$\min^t_u = \begin{cases} R_u(t) & \text{if } (s^t_u = 0 \& s^{t-1}_u = 1) \text{ or } R_u(t) < \min^{t-1}_u, \\ \min^{t-1}_u & \text{otherwise.} \end{cases} \quad (7)$$

$$\max^t_v = \begin{cases} R_v(t) & \text{if } (s^t_v = 1 \& s^{t-1}_v = 0) \text{ or } R_v(t) > \max^{t-1}_v, \\ \max^{t-1}_v & \text{otherwise.} \end{cases} \quad (8)$$

In Algorithm 1, for each edge $e$ in the stream, MONLAD will create and initialize some vertex-specific variables via INITIALIZE for the newcomers if $u$ or $v$ never appear, then it performs update with the above rules, and finally outputs the statistical features of each node at the current time, i.e., $B_*(t)$ and $F_*(t)$.

Figure 3 illustrates a running example of MONLAD. The state $s^t_u$ changes to 0 at $t_5$ and $t_{11}$ due to reaching a balanced state, $\max^t_u$ changes at $t \in \{t_2, t_4, t_6, t_9, t_{10}\}$, and $\min^t_u$ changes at $t_5$ and $t_{11}$.

*4.2.1 MONLAD with sliding window: MONLAD-W.* In real applications, transaction history involves a variety of accounts with different behaviors, which can be personal or corporate; fraudsters will avoid trading too frequently to evade detection, which differs from the behavior of company accounts that are used for frequent settlement or lopende rekening, so the global statistical features in MONLAD may be inappropriate to distinguish them. Furthermore, entirely accumulative counting also can result in some false positive detection, since normal accounts can also achieve balanced states periodically due to the credit card payback or fund transfers after payroll, thus we should rather capture the local behavior of a short period of time. So, we propose MONLAD-W, as the variant of MONLAD with a sliding time window, to compute the local features of B and F. We can use various statistics of the features w.r.t windows, i.e., max, mean. The detail of MONLAD-W algorithm is given in the supplement.

**Algorithm 2** ANOSCORE: Anomaly scoring with EVT

**Input:** Accounts features $B^+_*$ and $F'_* = (F_* - B_*)$; threshold $p$ and percentile $\alpha$ for fitting the PARETO distribution.
**Output:** Suspicious vertex set $\mathcal{M}$
1: $\mathcal{M} \leftarrow \{\emptyset\}$
   ▷ *Use the IQR and $Q_3$ to find the truncation thresholds*
2: Compute $IQR$s and $Q_3$s for $B^+_*$ and $F'_*$ to get the $b_1$ and $f_1$
   ▷ *Compute thresholds for the upper tails part of the distributions*
3: $b_2$ = PARETO($p, \alpha, \{B^+_u | F'_u = f_1; u \in \mathcal{V}\}$)
4: $f_2$ = PARETO($p, \alpha, \{F'_u | B^+_u = b_1; u \in \mathcal{V}\}$)
5: **for** $b \in [1, b_1]$ **do**    ▷ *Anomalies in Part I.*
6:    $\bar{f} \leftarrow$ PARETO($p, \alpha, \{F'_u | B^+_u = b\}$)
7:    $\mathcal{M} = \mathcal{M} \cup \{u | B^+_u = b \& F'_u > \bar{f}\}$
8: **for** $u \in \mathcal{V}$ **do**    ▷ *Anomalies in Part II.*
9:    **if** $(B^+_u > b_2 \& F'_u > f_1)$ or $(B^+_u > b_1 \& F'_u > f_2)$ **then** $\mathcal{M} = \mathcal{M} \cup \{u\}$
10: **for** $f \in [0, f_1]$ **do**    ▷ *Anomalies in Part III.*
11:    $\bar{b} \leftarrow$ PARETO($p, \alpha, \{B^+_u | F'_u = f\}$)
12:    $\mathcal{M} = \mathcal{M} \cup \{u | B^+_u > \bar{b} \& F'_u = f\}$
13: **return** $\mathcal{M}$

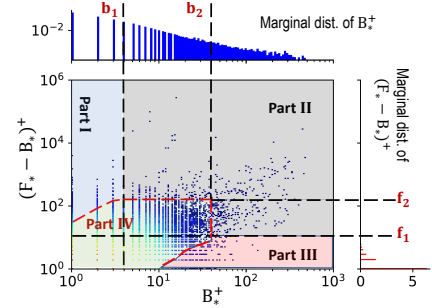

**Figure 4: The heatmap and marginal distributions of the statistical features from CBank and the example for ANOSCORE algorithm. $B^+_*$ and $F'_* = (F_* - B_*)$ are used, and the result at the end of the observation time are shown. ANOSCORE detects outliers in Part I, II, III, while the vertices in the high-density region (Part IV) are regarded as normal.**

## 4.3 ANOSCORE: Anomaly Scoring

Based on the above features of each account, we design the scoring schema, ANOSCORE, to measure their suspiciousness in the money laundering scenario, respond to the customer queries at any time $t$, and report a group of most suspicious agent accounts as $\mathcal{M}$.

The **Generalized Pareto Distribution** (GPD) is a 3-parameter distribution and has been used to find the law of extreme events (tails) within extreme value theory (EVT) [5] (Ref. the supplement)

The second theorem of EVT is the Pickands-Balkema-de Haan theorem [2, 25]. It states that given a random variable $X$, let $m \in \mathbb{R}$ and define a new random variable $X_m$, that intuitively represents the tail of $X$ past threshold $m$; define $F_m$ as the distribution of $X - m$ conditioned on $X > m$. Then, the conditional excess at threshold $m$ has CDF $F_m(x) = \mathbb{P}(X - m \leq x | X > m)$.

As the traits summarized in Sec. 3 and description in Sec. 4.1 show, the behavior of the agent accounts usually has great difference with the normal, this justifies using GPD to model the upper tail of the distribution of their features. The empirical result from the real data as shown in Figure 4 also verifies our motivation, i.e, the distributions of $B_*^+$ and $F_*' = F_* - B_*$ are highly skewed[5].

With the arrival of edges of a stream, MonLAD outputs $B_u(t)$ and $F_u(t)$ for all seen vertices $u \in \mathcal{V}$ at time $t$ (MonLAD-W is also suitable). So, we design AnoScore in Alg. 2 to answer the customer's query for anomalies in the stream at any time. AnoScore takes the positive features $B_*^+$ and $F_*' = F_* - B_*$, and the probability threshold $p$ and percentile $\alpha$ for Pareto distribution as input, finally returns the subset $\mathcal{M}$ containing the most suspicious vertices.

The function Pareto uses the upper-tails part exceeding at threshold $m$, corresponding to samples at the bottom $1 - \alpha$ of the ordered population, to fit its parameters, and then outputs the minimum value of fitted data $x$ such that it satisfies $\mathbb{P}(X - m > x | X > m) = 1 - F_m(x) < p$ for some parameter $p$ and $\alpha$, thus $p$ plays a similar role as the P-value to test the null hypothesis (i.e., the points are normal). we set $\alpha = 98\%$ in experiments as [36].

Based on the correlation between the features as Figure 4 shows, AnoScore detects those anomalies in three parts over the space of the joint distribution. Specifically, the decision boundaries $b_1$ and $f_1$ are determined by $Q_3 + K * IQR$ with $IQR = Q_3 - Q_1$[6], $b_2$ and $f_2$ are determined by the Pareto which takes in the samples $\{u\}$s with $F_u' = f_1$ and $B_u^+ = b_1$ respectively; we set $K = 1.5$ which is typically used for normally distributed data [13]. Then, the outliers in Part I, II, III are detected one by one (Lines 5 - 12), while Part IV is regarded as the normal. The red dashed line in Figure 4 highlights the final decision boundary for detecting anomalies. In general, $b_2 > b_1$ and $f_2 > f_1$ are always true for sufficient data from large-scale stream; we will set $b_2 = b_1$ and $f_2 = f_1$ for some possible situation where $b_2 \leq b_1$ and/or $f_2 \leq f_1$ due to the ill-posed data distribution.

THEOREM 3 (TIME COMPLEXITY[7]). *The time complexity of the MonLAD algorithm is linear with the number of edges in the stream, $O(|\mathcal{E}|)$; MonLAD-W uses $O(\frac{|\mathcal{T}|}{s} \cdot |\mathcal{V}_t|)$ time to compute the window feature in addition. AnoScore scales linearly with the number of the seen vertices, $O(|\mathcal{V}_t|)$, at current time $t$.*

## 5 EXPERIMENTS

We design experiments to answer the following questions:
(1) **Q1. Effectiveness for pattern detection:** How accurately does MonLAD [8] detect expected behavior patterns (**P1-P3**) of agent accounts for money laundering? Does AnoScore have advantage over other baselines?
(2) **Q2. MonLAD spots real-world agent accounts:** What patterns does MonLAD(-W) detect in real-world datasets? How about the behavior of the suspicious agent accounts?
(3) **Q3. Scalability:** Does our method scale linearly with the number of edges?

**Datasets:** We use two real-world datasets, the 'CBank' dataset from an anonymous bank under an NDA agreement; the Czech

---

[5]We use the difference feature since $F_* \geq B_*$ always hold for any vertex ('*').
[6]$Q_1$ and $Q_3$ represent 1st and 3rd quartile of samples respectively.
[7]The detailed proof of the Theorem 3 is given in the supplement.
[8]We use MonLAD(-W) to refer to MonLAD(-W) followed with AnoScore here.

Table 3: Statistics of Real-World Datasets.

| Dataset | # of Nodes | # of Edges | Time Span ($\mathcal{T}$) |
|---|---|---|---|
| CBank | 8.77 M | 47.44 M | Aug.7 - Aug.13, 2017 |
| CFD [21] | 11.37 K | 273.51 K | Jan.1,1993 - Dec.1,1998 |

Financial dataset ('CFD') is an anonymous transferring transaction data of Czech bank released for Discovery Challenge in [21]. Table 3 lists the statistical information of the two datasets.

**Baselines:** We select following methods as baselines, Spotlight [8], Midas-R [3], and SedanSpot [7] that detect anomalies in edge streams, and D-Cube [33], a batch yet fast algorithm that outputs close or better accuracy than its streaming version, DenseAlert [34].

**Evaluation Metric:** We evaluate the performance with *F1 score*. For the baselines detecting suspicious edges rather than vertices, we treat an edge as a hit if any of its ends are labeled anomalous. For a fair comparison, we report the best F1 score for those baselines returning a rank of suspicious edges.

**Experimental Setup:** All experiments are carried out on a 2.7GHZ Intel Xeon E7-8837 CPUs processor and 512GB RAM running Linux. MonLAD is implemented in Python. In all the experiments, we set $\alpha = 98\%$, $p = 0.05$ and $\delta_{up} = \delta_{down} = \epsilon = 10k$ for our methods (unless specified otherwise). We average the results over 5 trials for all synthetic experiments for anomalies injection.

### 5.1 Q1. Effectiveness in pattern detection.

#### 5.1.1 Effectiveness of MonLAD.
Considering the various behavior patterns in money laundering scenario as summarized in Sec. 4.1, we adopt numerous injection schema upon the CBank dataset to verify MonLAD's performance for spotting different kinds of anomalies. Based on the characteristics of patterns **P1-P3**, we control the behavior of agent accounts via different $f_*$ and $B_*$.

We use the *clean CBank* data, i.e., removing all suspicious accounts detected by MonLAD, as background for injection. For patterns **P1** and **P2**, we inject $|\mathcal{M}| = 200$ agent accounts as anomalies and randomly select the source or target accounts for money transfer-in and out; the amount of money for each transaction is sampled from a Dirichlet distribution [9] to ensure that the total amount (no more than $1E8$) is uniformly assigned to each edge; all the money transferred into an account is clear out and reaching a balanced state finally; the time of each transaction is randomly sampled from the time range.

Figure 5(a)-5(b) show the performance of MonLAD and baselines for detecting different patterns. We report the results of MonLAD with different $p$s (0.01 to 0.05 with step 0.01) for clear comparisons. As we can see, MonLAD consistently outperforms all baselines for **P2** pattern by achieving at least 68% improvement in Fig. 5(a) (with $p = 0.01$). Fig. 5(b) demonstrates the performance for detecting **P1** behavior pattern, we can see FlowScope also achieves high accuracy (triangle marked) since these found accounts have fewer transfer connections but higher average amounts. Compared with Fig. 5(a), however, FlowScope gives the worst accuracy for low-density cases. We compare the accuracy and speed with baselines

---

[9]It generates one sample (edge) each time with the parameter $\alpha = [\alpha_1, \ldots, \alpha_{|\mathcal{E}|}]$ with $\alpha_i = 100$ for $1 \leq i \leq |\mathcal{E}|$ to guarantee a small variance among the sampling.

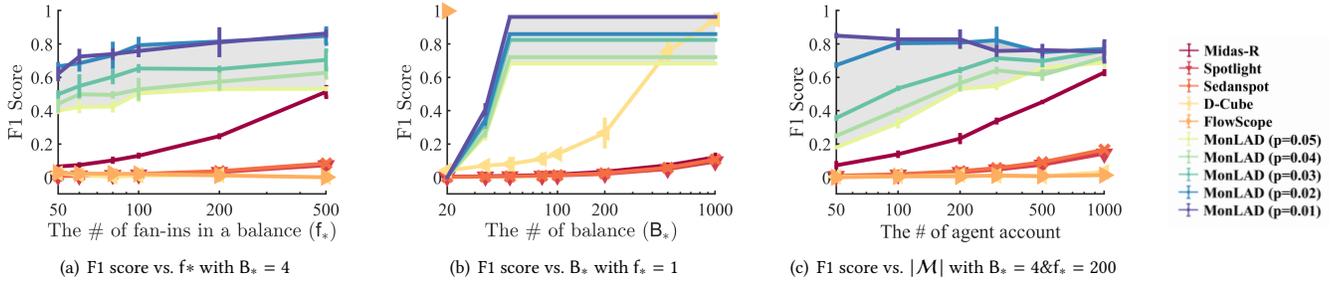

Figure 5: Performance comparison in F1 Score with standard deviation. Our method consistently outperforms baselines in detecting injected anomalies, the shadow area shows the performance range of MonLAD for $p$ varying from 0.01 to 0.05.

under two specific settings, $f_* = 500$ & $B_* = 4$ in Fig. 5(a) and $f_* = 1$ & $B_* = 1k$ in Fig. 5(b). Fig. 2(b) shows the result, where the empty and filled markers correspond to the above two cases; we can see that MonLAD achieves nearly 4× speedup over FlowScope with very similar accuracy and outperforms other competitors.

We evaluate the performance of MonLAD against the number of injected agent accounts ($|\mathcal{M}|$) and give the result in Fig. 5(c). MonLAD achieves the best F1 score and has obvious advantages over baselines; its accuracy decreases as $|\mathcal{M}|$ increases for $p \leq 0.02$, since the injection gradually dominates the tail part of data, resulting in lower recall when the precision has reached the maximum.

## 5.2 Q2. Spotting real-world agent accounts

*5.2.1* **Analysis of MonLAD.** We apply MonLAD to CBank data for detecting suspected real-world agent accounts and analyze their behavior patterns. As Figure 2(a) shows, MonLAD spots various types of anomalies for money laundering where the decision boundary is denoted as the red dashed line. In the absence of labels, we randomly select some detected accounts from different parts for case study about their transfer structures, residual series, and statistical features ($B_*$ & $F_*$). Moreover, 12 accounts are manually verified to be fraudulent agents, they can be easily and accurately caught by MonLAD and are labeled with 'criminal-icon' near $p_2$ in Fig.2(a).

Figure 1 and Figure 6 show the behavior patterns of some selected accounts ($p_1 - p_4$). For the account $p_3$ in Fig. 1, it receives 63 transfers from two different accounts ($\approx 34.72M$ in total) and transferred them to 6 other accounts multiple times, there are two accounts accounted for the majority and other small amount transfers are very likely to be camouflage transactions; from the residual series, we can see that a large amount of money flowed through $p_3$ with **P1** behavior leading to reaching a number of balanced states, which is a typical suspicious pattern.

Additionally, from Figure 6(a)-6(b), we find that the account $p_1$ has similar transfer structure with $p_3$, but it has multiple fan-ins with the behavior similar as **P2** pattern as Fig. 6(c) shows; so $p_1$ transfers money in a mixed manner of **P1** and **P2**, i.e. **P3**. As we observe from Fig. 6(d)-6(f), the fan-ins of the verified account $p_2$ is more than 1,000× its fan-outs, which is consistent with the typical **P2** pattern. For the account $p_4$, it has the same behaviors as **P2** based on the zoom-in sub-figure in Fig. 9(b) and Fig. 6(i), while it is less suspicious than $p_2$ due to close to the decision boundary.

In contrast, the residual of the normal accounts in Part IV rarely reaches balanced states, we provide detailed analysis for the behavior of some randomly selected normal examples in the supplementary. Therefore, MonLAD indeed finds some suspicious agent accounts for money laundering.

*5.2.2* **Analysis of MonLAD-W.** We verify the effectiveness of MonLAD-W for detecting suspicious behavior with periodical balanced pattern and compare with MonLAD. We use CBank dataset and set the parameter of MonLAD same as the Sec. 5.2. We set window size $K = 1h$, sliding stride $s = 1h$ for MonLAD-W which use the maximum value of the statistical feature series as score, $p =$1e-3 for AnoScore.

Figure 7(a) illustrates residual series of two detected accounts as examples for the case study. The account at the bottom is detected by MonLAD-W but not MonLAD. It achieves high-frequent balanced states within one hour (each takes less than $5min$), which is rather suspicious; it was active only on Aug. 4th, which results in a smaller value of B than the Bs of those normal accounts that have been active since the early days from the point of cumulative counting. Therefore, the anomalous accounts like above mentioned one will not be detected by MonLAD. Furthermore, we randomly select another account that has the same features (B = 8 & F′ = 1) from MonLAD as the above account. As shown in the top of Fig. 7(a), we can see that it had a similar periodical balanced pattern, yet with a much longer period (i.e. $\approx 0.5\ d$). Although the top one is more likely to be a normal account, MonLAD cannot distinguish it from the more suspicious one at the bottom. In contrast, MonLAD-W is able to avoid such false positive (FP) detection since it only counts the number of features within a specified time window. For MonLAD-W, the feature values of these accounts are B = 1&F′ = 0 (top) and B = 7&F′ = 1 (bottom).

Therefore, MonLAD-W is able to capture the local behavior pattern while avoiding some FN and FP cases caused by the accumulative effect in MonLAD.

*5.2.3* **Development at a real-world platform.** We also apply MonLAD to a real money transfer dataset collected from Tencent's WeChat App, which consists of millions of accounts and transaction records for one week [10] with the format (source, destination, time, amount).

---
[10] Concrete information is not public due to privacy protection.

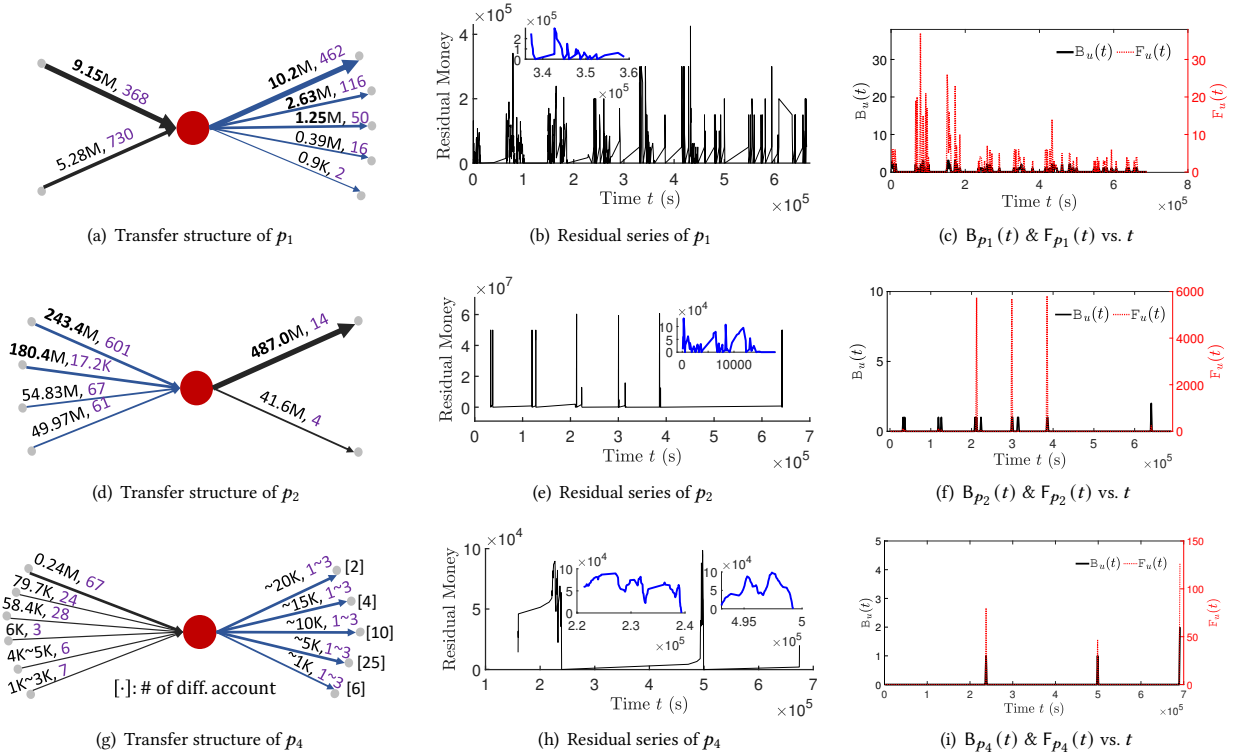

**Figure 6: Case study of the detected anomalies in real-world data (CBank). Here illustrate the transfer structure, residual series $R_*(t)$, and $B_*(t)$ & $F_*(t)$ of those randomly selected suspicious accounts ($p_1, p_2, p_4$) labeled in Figure 2(a).**

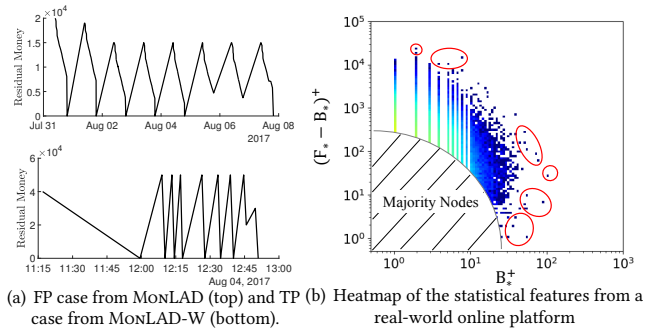

(a) FP case from MonLAD (top) and TP case from MonLAD-W (bottom).   (b) Heatmap of the statistical features from a real-world online platform

**Figure 7: Performance for real data and case study. (a) Two example detection cases with periodical pattern as the FP (false positive) from MonLAD and the TP (true positive) from MonLAD-W. (b) The heatmap of the features from MonLAD based on the real data from Tencent Wechat App and some manually verified suspected fraudster accounts.**

Figure 7(b) shows the heatmap of the statistical features. Due to the limitation of accessible data, we only randomly selected the most suspicious 30 accounts that are marked with red circles from the detection parts in Fig. 7(b), and manually verified according to the profile information, they turn out to be some suspected money laundering accounts (we use $p$ =1e-3 for AnoScore here).

### 5.3 Q3. Scalability

We measure how rapidly MonLAD's update time increases as the stream grows. We used the accumulative edge streams of CBank per 12 hour and gives the running time of the algorithm until the specific time. Figure 2(c) shows the linear scaling of MonLAD's running time with the number of edges of the stream.

## 6 CONCLUSIONS

In this paper, we propose MonLAD and MonLAD-W for detecting the money laundering anomaly agent accounts in a transaction stream; we introduce a statistical score schema, AnoScore, to find anomalies that have obvious deviation behavior. The experiments on real-world datasets show that our method MonLAD achieves state-of-the-art performance and interpretable results, and our methods are also linearly scalable.

## ACKNOWLEDGMENTS

This paper is partially supported by the National Science Foundation of China under Grant No.91746301, 61772498, U1911401, 61872206. This paper is also supported by the Strategic Priority Research Program of the Chinese Academy of Sciences, Grant No. XDA19020400 and 2020 Tencent Wechat Rhino-Bird Focused Research Program.

# A APPENDIX

## A.1 Effectiveness of AnoScore.

Given the statistical features ($B_*$ and $F_*$), how does the AnoScore algorithm perform? Here, we verify the effectiveness of AnoScore by directly injecting various anomalies into the feature space and compare its detection performance with some classic outlier detection approaches.

In this experiment, we use CFD dataset and inject anomalies into three different parts in the $F'_*$ vs. $B_*$ space like Figure 2(a). We choose 6 methods implemented in PyOD [45] as baselines, KNN, PCA, Clustering-Based Local Outlier Factor (CBLOF), Histogram-based Outlier Score (HBOS), Isolation Forest (IF), Minimum Covariance Determinant (MCD). Here, we adopt *FAUC* (the Areas Under the Curve of F1 metric) [17] to measure the performance and normalize $B_*$ or $F'_*$ in horizontal axis to make FAUC in [0, 1]; the higher FAUC indicates better performance. We set $\alpha = 0.5$ and $p = 0.2$ for AnoScore due to the limited data size. The parameter setting of baselines and detailed injection schema are given in A.6.

Table 4: AnoScore outperforms other outlier-detection methods over CFD dataset

| Anomalies type | CBLOF | HBOS | IF | KNN | MCD | PCA | AnoScore |
|---|---|---|---|---|---|---|---|
| Part I ($F'_*$) | 0.706 | 0.202 | 0.494 | 0.131 | 0.188 | 0.573 | **0.741** |
| Part III ($B_*$) | 0.661 | 0.324 | 0.775 | 0.554 | **0.839** | 0.590 | 0.687 |
| All Parts ($F'_*$) | 0.791 | 0.612 | 0.845 | 0.570 | 0.750 | **0.848** | 0.810 |
| All Parts ($B_*$) | **0.808** | 0.608 | 0.776 | 0.556 | 0.733 | 0.752 | 0.736 |
| Average | 0.742 | 0.437 | 0.722 | 0.453 | 0.627 | 0.691 | **0.744** |

Table 4 shows the detection results of different detection methods for various injection settings. On average, AnoScore achieves the best performance and is in par with CBLOF and IF, they have own strength for different parts which corresponds to different behavior patterns. However, AnoScore is more interpretable and intuitive due to unitizing the statistical deviations for the money laundering scenario. Besides, we can see that some method also performs well in some specific parts, e.g, MCD for 'Part III' and PCA for 'All Parts ($F'_*$)'; HBOS and KNN have the worst performance. In summary, our MonLAD is more flexible to in favor of different outlier detection methods and achieves appealing results.

## A.2 MonLAD-W algorithm

Algorithm 3 describes the framework of MonLAD-W, which can seamless connect with MonLAD, it computes and outputs the local statistic features of the current time window with Line 4-14.

As can be seen, MonLAD-W yields $B_{win}$ and $F_{win}$ for all accounts, forming a group of series, across the time window; thus we can also analyze trends or statistics for them, like maximum, mean, variance, etc. The maximum value is used in our experiments.

## A.3 Generalized Pareto Distribution

Specified by the parameters location $\mu$, scale $\sigma > 0$, and shape $\xi$, the cumulative distribution function (CDF) of GPD is defined as,

$$\text{GPD}_{\mu,\sigma(t),\xi}(x) = \begin{cases} 1 - \left(1 + \frac{\xi(x-\mu)}{\sigma}\right)^{-1/\xi} & \text{if } \xi \neq 0 \\ 1 - \exp\left(-\frac{x-\mu}{\sigma}\right) & \text{if } \xi = 0 \end{cases}$$

The second theorem of EVT is the Pickands-Balkema-de Haan theorem [2, 25]. It states that given a random variable $X$, let $m \in \mathbb{R}$ and define a new random variable $X_m$, that intuitively represents the tail of $X$ past threshold $m$; define $F_m$ as the distribution of $X - m$ conditioned on $X > m$. Then, the conditional excess at threshold $m$ has CDF $F_m(x) = \mathbb{P}(X - m \leq x | X > m)$.

Being agnostic to the distribution of original data under a weak condition, the key properties of GPD are its flexibility in smoothly interpolating between light- and heavy-tailed regimes (extreme events), and its *universality* property.

PROPERTY 1 (UNIVERSALITY OF GPD). *Let $F$ be any distribution function from a broad class of distributions[11]. For the CDF of the GPD, there exists $\xi$ and $\sigma(m)$ that approximate the tail of $F$ arbitrarily closely, that is, $\lim_{m \to m_{\max}} \sup_x |F_m(x) - \text{GPD}_{0,\sigma(m),\xi}(x)| = 0$, where $m_{\max}$ is the right endpoint of $F$ and can be $\infty$.*

---

**Algorithm 3** MonLAD-W: MonLAD with sliding time window

**Input:** Stream of edges over time; thresholds $\delta_{up}, \delta_{down}$, and $\epsilon$; window size $K$, sliding stride $s$.
**Output:** Statistical features per window.

1: $t_{begin} = t_{end} = 0$;
2: Initialize feature vectors $\mathbf{B}_{t_{begin}}(*) \leftarrow 0$, $\mathbf{F}_{t_{begin}}(*) \leftarrow 0$
3: **while** new edge $e = (u, v, w_{u,v}, t)$ is received **do**
4:     **if** $t > t_{end}$ **then**    ▷ *Receive an unseen time tick*
5:         $\mathbf{B}_{t_{end}}(*) \leftarrow \mathbf{B}(*), \mathbf{F}_{t_{end}}(*) \leftarrow \mathbf{F}(*)$   ▷ *Record for $t_{end}$*
6:         $\bar{t} = K + s * \lfloor \frac{t-K}{s} \rfloor$    ▷ *Bound of current window*
7:         **if** $t_{end} \geq K$ & $t_{end} \leq \bar{t}$ **then**    ▷ *Window update*
            ▷ *Compute feature vectors w.r.t. the current window*
8:             $\mathsf{B}_{win} = \mathbf{B}_{t_{end}}(*) - \mathbf{B}_{t_{begin}}(*)$
9:             $\mathsf{F}_{win} = \mathbf{F}_{t_{end}}(*) - \mathbf{F}_{t_{begin}}(*)$
10:           $t_{begin} \leftarrow \max_i \{t_i | t_i \leq t_{begin} + s\}$
11:           **for** $t_i < t_{begin}$ **do**    ▷ *Window slides with $s$*
12:             Remove $\mathbf{B}_{t_i}(*)$ and $\mathbf{F}_{t_i}(*)$
13:           **yield** $\mathsf{B}_{win}, \mathsf{F}_{win}$    ▷ *Output the current features*
14:         $t_{end} \leftarrow t$
15:     $\mathbf{B}(u), \mathbf{F}(u), \mathbf{B}(v), \mathbf{F}(v) = \text{MonLAD}(e, \delta_{up}, \delta_{down}, \epsilon)$
    **return**

---

## A.4 Analysis of Time Complexity

For MonLAD, for each edge, the algorithm will update the corresponding variables according to the rules, which takes $O(1)$. Thus, the time complexity of the MonLAD algorithm is linear with the number of edges in the stream, $O(|\mathcal{E}|)$.

For MonLAD-W, additionally, the most time consuming steps are Lines 8 − 13 in Algorithm 3, which are executed at the end of each window. Let $|\mathcal{T}|$ be the number of time ticks, $K$ be the window size and $s$ be the sliding stride size, then the number of time windows will be $\frac{|\mathcal{T}|-K}{s}+1$; vectors subtraction at line 8−9 takes $O(|\mathcal{V}_t|)$. Therefore, the additional time complexity of MonLAD-W is $O(\frac{|\mathcal{T}|}{s} * |\mathcal{V}_t|)$.

---
[11]This class includes almost all commonly used distributions [6].

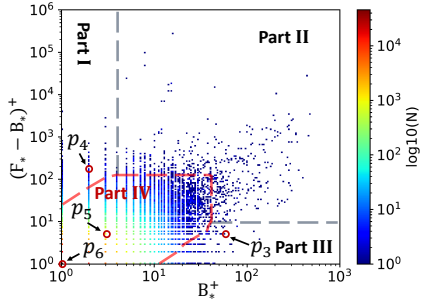

Figure 8: Heatmap of the statistical features for CBank. Here we randomly pick 2 normal accounts $p_5$ and $p_6$ ($\mathsf{B}_{p_6} = \mathsf{F}'_{p_6} = 0$) from the Part IV.

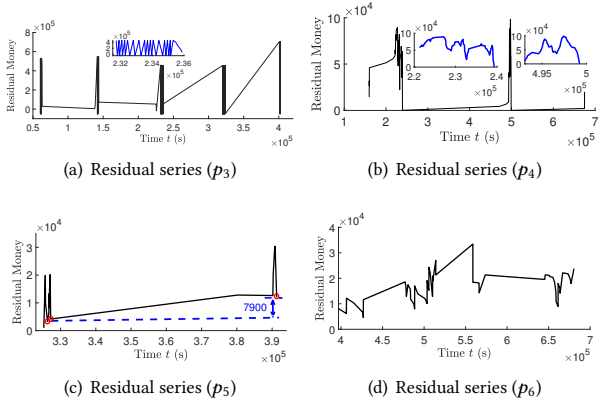

(a) Residual series ($p_3$)

(b) Residual series ($p_4$)

(c) Residual series ($p_5$)

(d) Residual series ($p_6$)

Figure 9: The time series of the residual money of several suspicious accounts and other normal accounts randomly sampled from CBank dataset (the parameters for MonLAD are $\delta_{up} = \delta_{down} = \epsilon = 10\mathbf{k}$).

For AnoScore, the thresholds computing at Lines 3 − 4, Line 6, and Line 11 for the upper tails part of distribution will take $O(1)$, and it takes $O(\mathcal{V}_t)$ to determine whether each node is abnormal in Lines 8 − 9. Thus, AnoScore scales linearly with the number of the seen vertices, $O(|\mathcal{V}_t|)$, at the current time $t$.

### A.5 Normal Behavior Patterns.

To further verify the behavior patterns detected with MonLAD, we randomly select 2 normal accounts (fall in the part IV of the features heatmap in Figure 8) from CBank dataset as comparison.

The results are illustrated in Figure 9. In Figure 9(c), we find that the account $p_5$ has the same behaviors as **P1** pattern (i.e., one-time fan-in and immediate fan-out), but it reaches the balanced state only three times, which are labeled with red circle marker, and the total amount of transfers is far less than the suspicious accounts $p_3$ and $p_4$ analyzed in the main paper (see Figure 9(a)-9(b)), thus $p_5$ is more likely to be a normal account. Meanwhile, as shown in Figure 9(d), there is no obvious pattern in the residual series of the account $p_6$ ($\mathsf{B}_{p_6} = \mathsf{F}'_{p_6} = 0$) and it never reaches a balanced state, which is consistent with the normal behavior.

In summary, those results suggest that the normal accounts rarely reach a balanced state compared with the anomaly accounts whose behavior are more suspected, MonLAD can spot the anomalies based on those behavior patterns.

### A.6 Detailed Experiment Results.

Here we provide the detailed information about the experimental settings (baseline parameter settings and injection scheme), qualitative analysis and comparison for the additional results of AnoScore and baselines.

**Implementations.** In this experiment, we consider 6 outlier detection algorithms implemented by [45] for comparison:

- *Clustering-Based Local Outlier Factor (CBLOF)* [12]: We set the number of clusters $k = 8$, the coefficients for deciding small and large clusters $\alpha = 0.9$ and $\beta = 5$, as recommended in the original work.
- *Histogram-based Outlier Score (HBOS)* [10]: We set the number of bins $k = 10$.
- *Isolation Forest (IF)* [18]. We set the number of trees to $t = 100$ and the sub-sampling size to $\psi = 256$, as recommended in the original work.
- *KNN* [26]: We set the number of nearest neighbors $k = 5$.
- *Minimum Covariance Determinant (MCD)* [11, 31]: We set the proportion of points to be included in the support of the raw estimate $h_j = 0.8$, leading to a better accuracy than the setting in the original work.
- *PCA* [35]. We use euclidean distance as distance metric and keep all principal components to calculate the outlier scores.

**Injection scheme.** We use the clean CFD dataset as the base for injection after removing all suspicious accounts detected by MonLAD with the corresponding detection thresholds are $b_1 = 13$, $b_2 = 20$, $f_1 = 95$ and $f_2 = 120$.

For each part, we inject 20 points as anomalies. Specifically, for the Part I, we vary $\mathsf{F}'_*$ from 100 to 130 and randomly select $\mathsf{B}_*$ from 1 to the threshold $b_1$ (= 13) and ensure injected anomalies fall in this desired part. Similarly, for the Part III, we vary $\mathsf{B}_*$ from 15 to 35 and randomly select $\mathsf{F}'_*$ from 1 to the threshold $f_1$ (= 95). For the mixture cases, we change the above two experimental features (i.e., $\mathsf{F}'_*$ and $\mathsf{B}_*$) separately and randomly inject 40 points into Part II. And when one experimental feature changes, such as $\mathsf{F}'_*$ (or $\mathsf{B}_*$), we will fix $\mathsf{B}_*$ (or $\mathsf{F}'_*$) of Part III (or Part I) to the threshold $b_2$ (= 20) (or $f_2$ (= 120)).

**Analysis and Comparison.** Based on the above performances, we analyze each method as follows: HBOS assumes the feature independence while $\mathsf{F}'_*$ and $\mathsf{B}_*$ are related; KNN ranks points by the distance to its $k$ nearest neighbor, however, we assume that the suspicious accounts usually locate on the upper tail of data so that close to the normal. Additionally, MCD is proposed based on Gaussian-distributed data while the distribution of our features are highly skewed as shown in Fig. 3 in the main paper. In contrast, methods without special assumptions perform better. CBLOF focuses on the physical significance of outliers, PCA utilizes the correlation between data and IF focus on isolated points, yet they are designed for general outliers detection purpose and are not fully suitable for the money laundering detection scenario. However, our proposed AnoScore can avoid the above problems and achieves higher accuracy.